\documentclass[]{spie}  %>>> use for US letter paper

\usepackage{amsmath,amsfonts,amssymb}
\usepackage{graphicx}
\usepackage[colorlinks=true, allcolors=blue]{hyperref}
\usepackage{dcolumn}% Align table columns on decimal point
\usepackage{bm}% bold math
\usepackage{tikz} % for tikz picture
\usepackage[utf8]{inputenc}
\usepackage{braket}    % Allows Dirac Notation
\usepackage{subfig}    % Allows for subfigures
\usepackage{comment}   % Allows for commenting out large sections
\usepackage{float}	   % Allows for placing figures more precisely ? 
\usepackage{multirow}
\usepackage{color}
\usepackage{fancyhdr}
\usepackage{tabularx}
\usepackage{setspace}
\usepackage[normalem]{ulem} % Dash added this for strikethrough support
\usepackage[export]{adjustbox}
\usepackage{stfloats}%The stfloats package also ameliorates the situation somewhat, and makes LaTeX honour “[b]” placement as well.
\usepackage{tabu}
\usepackage{caption}
\usepackage{array} % Allows for tabular {m} command
\usepackage[colorinlistoftodos,prependcaption]{todonotes} % Dash added this for the todo notes.

\newcommand{\be}[1]{\begin{equation}\label{#1}}
\newcommand{\ee}{\end{equation}}
\newcommand{\bea}[1]{\begin{eqnarray}\label{#1}}
\newcommand{\eea}{\end{eqnarray}}

\newcommand{\Fig}[1]{Fig.(\ref{#1})}

\newcommand{\Eq}[1]{Eq.(\ref{#1})}
\newcommand{\App}[1]{Appendix~\ref{#1}}
\newcommand{\Sec}[1]{Section~\ref{#1}}
\title{Backscattering and Hong-Ou-Mandel Manifolds \\in Microring Resonators}

\author[a,b]{Peter L. Kaulfuss}
\author[a]{Paul M. Alsing}
\author[a,b]{Richard J. Birrittella}
\author[c]{Dashiell L.P. Vitullo}
\affil[a]{Air Force Research Laboratory, Information Directorate, 525 Brooks Rd, Rome, NY, 13411, USA}
\affil[b]{Griffiss Institute, 592 Hangar Rd STE 200, Rome, NY 13441, USA}
\affil[c]{DEVCOM Army Research Laboratory, 2800 Powder Mill Road, Adelphi, MD 20783, USA}

\authorinfo{corresponding author: Peter Kaulfuss. E-mail:  plk3729@rit.edu}

\begin{document} 
% put the page number on the RIGHT or LEFT
%\fancyhead[R]{\ifnum\value{page}<2\relax\else\thepage\fi}

\maketitle

\begin{abstract}

We investigate the effect of backscattering on the Hong-Ou-Mandel manifold (HOMM) that manifests in double-bus mircoring
resonators (MRRs). The HOMM represents higher-dimensional parameter solutions for the complete destructive interference
of coincident detection in the HOM effect. To model the backscattering, we introduce a set of internal `beam splitters' inside the ring that allow photons to `reflect' into new counter-propagating modes inside the MRR. We find that the one-dimensional HOMM in MRRs investigated here is extremely robust against deterioration due to backscattering, even in a linear chain of identical MRRs. Further we find that a small amount of backscattering introduced into a chain of non-identical MRRs connected in parallel could be desirable, causing them to behave more like chain of identical MRRs.
\end{abstract}

% Include a list of keywords after the abstract 

\section{Introduction}
It has been shown \cite{Hach:2014,Alsing_Hach:2017a,Kaulfuss_thesis:2021,Kaulfuss:2023.Identical,eHOM_Alsing:2022,eHOM_Conf_Alsing:2022}, that the microring resonator (MRR) provides an enhancement to the parameter space for achieving the Hong-Ou-Mandel (HOM) effect \cite{HOM:1987} in comparison to a simple beam splitter (BS) or directional coupler. A BS can only achieve the HOM destructive interference effect for coincident outputs at a single operational point (a balanced 50/50 BS). The higher dimensional space present in MRRs provides an entire manifold of solutions for tunable parameters that yield the HOM effect. We will refer to this manifold of solutions as the Hong-Ou-Mandel Manifold (HOMM). The HOM effect continues to be a fundamental tool in a variety of quantum applications \cite{Bouchard_2021} and this extension allows the HOM effect to be achieved more readily for a large range of tunable MRR parameters.

The double-bus MRR (or add-drop MRR) can be thought of as taking a beam splitter (single parameter) and expanding it into (in general) three parameters: two directional couplers (BS) and an internal phase angle (somewhat analogous to a Mach-Zehnder Interferometer)\footnote{The relationship between an MRR and a conventional beam splitter is discussed in detail in Appendix C of Kaulfuss, et al. (2022)\cite{Kaulfuss:2023.Identical}}. These three parameters are coupled together to form higher dimensional manifolds in which the HOM destructive interference condition can be met, which we denote as a HOMM. We simplify the MRR by allowing the two directional couplers to have the same coupling parameters (identical). As a result, we reduce our MRR to two tunable parameters, $\tau$ and $\theta$. Here, $\tau$ represents the coupling between the MRR and the waveguides via both directional couplers, and $\theta$ is the round-trip phase shift of the MRR. In the case of the MRRs investigated in this paper, the HOMM is a one-dimensional curve, $\tau(\theta)$ \cite{Kaulfuss:2023.Identical}. The increased flexibility of MRRs in combination with the ability to achieve the HOM effect for a wide range of parameters (HOMM) makes the MRR a powerful tool in achieving the HOM effect. 

In this work, we investigate the HOMM in MRRs subject to backscattering which allows photons circulating in one direction, to back scatter into a counter circulating mode.
Backscattering is commonly investigated in MRRs using a temporal coupled mode theory \cite{MRRReview:Bogaerts:2012,BackscatterReview:Li:2016,Hance:2021,Ballesteros:2011,McCutcheon:2021,Little:1997} and backscattering in MRRs is modeled from two main sources: either (1) ``sidewall roughness'' inside the MRR or (2) backscattering at the directional couplers. In this paper we propose a simple model for backscattering due to the sidewall roughness inside an MRR and focus our results on the HOMM achieved in MRRs. We investigate the effect of backscattering on the HOMM and determine for what level of backscattering the HOMM in MRRs persists. It is easy to imagine that the introduction of backscattering in a resonant device could have a destructive effect, especially on something like the HOM effect which is inherently an interference effect.  

The overview of the paper is as follows: in \Sec{Theory} we describe the theory of our backscattering MRR model. We define new modes and set up boundary condition equations that lead to our overall transfer matrix for the MRR, which we then use to define the HOMM condition. In \Sec{SingleMRR} we show the results on the HOMM for a single MRR with backscattering. In \Sec{otherinout} we investigate some input/output combinations other than our standard $\text{AB}_{\text{out}}\text{AB}_{\text{in}}$ to see if it is possible to obtain the HOMM on backscattered modes. In \Sec{Parallel} we show the results on the HOMM for parallel linear chains of identical MRRs with backscattering. In \Sec{Nonidentical} we show the results on the HOMM for parallel linear chains of non-identical MRRs and highlight the upside of a small amount of backscattering in non-identical MRR chains. Lastly, in \Sec{Conclusion}, we conclude that the HOMM in MRRs is extremely robust against backscattering because the optimal operating region for the HOMM in MRRs is off resonance. In \App{Appendix:1} we provide a helpful mnemonic for interpreting interference effects in the input/output MRR transfer matrix. 

%In \App{Appendix:2} we provide additional plots which include scattering loss along the MRR.

\section{Theory} \label{Theory}
\begin{figure}[H]
    \centering
    \begin{tikzpicture}[scale=1.5]
        \input{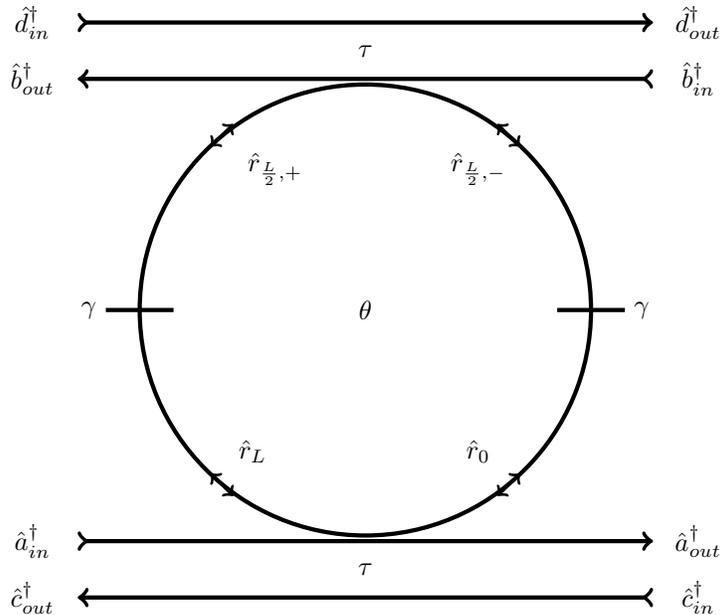}
    \end{tikzpicture}
    \caption{Diagram of a double-bus MRR with backscattering.} 
    \label{back reflection MRR diagram}
\end{figure}

We model backscattering where coupling arises between clockwise and counterclockwise propagating modes in a MMR, which can arise from imperfections in fabrication or dust accumulation in the evanescent field mode. As such, we must create new modes in the waveguides to allow for photons to travel in both circulating directions in each waveguide. We label the  inputs and outputs  for counter clockwise propagating modes as: $\hat{a}_{in}$, $\hat{a}_{out}$, $\hat{b}_{in}$, and $\hat{b}_{out}$. 
We now add $\hat{c}_{in}$, $\hat{c}_{out}$, $\hat{d}_{in}$, and $\hat{d}_{out}$ for the clockwise propagating modes,
where the $\hat{c}$ and $\hat{d}$ mode inputs are located at the $\hat{a}$ and $\hat{b}$ outputs, and the $\hat{c}$ and $\hat{d}$ mode outputs are located at the $\hat{a}$ and $\hat{b}$ inputs (see \Fig{back reflection MRR diagram}). We must also create a second set of interior modes to allow photons to also travel in the both directions within the MRR. We designate the interior modes of the MRR by $\hat{r}$, and employ four interior mode locations: $\hat{r}_0$, $\hat{r}_{\frac{L}{2},-}$, $\hat{r}_{\frac{L}{2},+}$, and $\hat{r}_L$. 
The interior modes are labeled by subscripts indicating their location (distance $z\in[0,L]$)
%$z=[0,L]$) % DASH: Is this change OK, Peter?
inside the MRR (measured in the counter clockwise direction) as shown 
in  \Fig{back reflection MRR diagram}. 
In order to relate the interior modes in our boundary conditions, we use a superscript to denote whether the mode is for a photon traveling in the clockwise (cw) or counterclockwise (ccw) direction. 
%
%===============
%INSERT NEW FIGURE 1
%===============
%==============================
%
%INSERT PARAGRAPH DESCRIBING ALL THE VARIABLES IN THE DIAGRAM
\noindent The interior modes are required so that at each junction we have a BS-type interaction with two modes entering and two modes exiting the BS. The modes $\hat{r}_0$ and $\hat{r}_L$ are the interior modes immediately inside and just before exiting the MRR for the counter clockwise modes $\hat{r}^{ccw}_{z}$. Similarly, $\hat{r}_{L/2,-}$ and $\hat{r}_{L/2,+}$ are the modes just inside and just before exiting the MRR for the clockwise circulating modes $\hat{r}^{cw}_{z}$. 
Phase accumulation within the MRR occurs, in general, as 
$\hat{r}_{z} = e^{i \theta (z-z_0)/L}\hat{r}_{z_0}$ for $z-z_0>0$ in both circulating modes. In the following, the interior modes will be eliminated in order to obtain the algebraic relationship between the input and output modes directly.

In general, $\tau$ and $\kappa$ are coupling parameters between the MRR and the waveguide via the directional coupler, where $\tau$ represents transmission and $\kappa$ represents reflection into/out of the MRR. We will assume that both directional couplers are identical. We choose $\tau$ to be a real valued parameter (without loss of generality \cite{tau:real:note})
%\textcolor{red}{[curious: if a coupling phase is added, does much change? Is this a `without loss of generality' situation?] Peter: Come back to this...} 
to represent the coupling, and 
to satisfy the reciprocity relations of the directional coupler, we take $\kappa$ to be 
\begin{equation}
    \kappa=\sqrt{1-\tau^2}.
\end{equation}
Thus, we define the unitary beam splitter (BS) transfer matrix between input to output modes as
\begin{equation}
\left(
\begin{array}{cc}
\tau & \kappa \\
-\kappa & \tau
\end{array}
\right). \label{BSmatrix}
\end{equation}
We have assumed that the transition elements at the directional couplers are the same regardless of which direction photons are traveling around the ring (i.e. the $\tau$ and $\kappa$ at the directional couplers are the same regardless of whether the photon is traveling in the counterclockwise or clockwise direction, the only difference is which mode it will output into). 

To model the backscattering, we assume that each photon has a chance to backscatter at every point inside the MRR. We condense these infinitely-many possible backscattering events into two effective internal beam splitters. Transmissions through these internal beam splitters will keep the photon circulating in the same direction and reflections cause the photon to circulate in the opposite direction. This is the only mechanism in this model for photons to change their propagation direction in the MRR. The clockwise and counterclockwise sets of interior MRR modes have different coupling relations at the directional couplers which determines the output modes that they will exit. 
Specifically, if a photon couples out of the MRR while traveling in the counterclockwise direction it will output into either $\hat{a}_{out}$ or $\hat{b}_{out}$. However, if a photon couples out of the MRR while traveling in the clockwise direction it will output into the $\hat{c}_{out}$ or $\hat{d}_{out}$ modes. 

We have modeled the backscattering as effective reflections at an internal beam splitter, where we have chosen $\gamma$ to be the real transmission value of the inner-ring beam splitter. This means that the value of $\gamma$ is the actual percentage of photons that continue to transmit around the ring in the same direction they started in (i.e. the corresponding equivalent  transmission and reflection coefficients for the interior backscattering  beam splitter would be $\sqrt{\gamma}$ and $\sqrt{1-\gamma}$). 
That is, the BS for the backscattering is given by the input/output unitary transfer matrix 
%
%\begin{equation}
$
\tiny{
\left(
\begin{array}{cc}
\sqrt{\gamma} & \sqrt{1-\gamma} \\
-\sqrt{1-\gamma} & \sqrt{\gamma}
\end{array}
\right)
}.
$
%\end{equation}
%
With this in mind the transmission/reflection (BS) equations in terms of the annihilation operators are as follows:
% \begin{align}
%     \hat{a}_{out}=\tau \, \hat{a}_{in} + \kappa \, \hat{r}_L^{ccw}, \\
%     \hat{r}_0^{ccw}=-\kappa \, \hat{a}_{in} + \tau \, \hat{r}_L^{ccw}, \\
%     \hat{b}_{out}= \tau \, \hat{b}_{in} + \kappa \, \hat{r}_{\frac{L}{2},-}^{ccw}, \\
%     \hat{r}_{\frac{L}{2},+}^{ccw}=-\kappa \, \hat{b}_{in} + \tau \, \hat{r}_{\frac{L}{2},-}^{ccw}.
% \end{align}

\begin{alignat}{3}
    \hat{a}_{out} &=\tau \, \hat{a}_{in} + \kappa \, \hat{r}_L^{ccw},  \;\;\;\;\;\;\;\;\;\;\;\;\; \hat{r}_0^{ccw} &&=-\kappa \, \hat{a}_{in} + \tau \, \hat{r}_L^{ccw}, \nonumber \\
    \hat{b}_{out} &= \tau \, \hat{b}_{in} + \kappa \, \hat{r}_{\frac{L}{2},-}^{ccw},\;\;\;\;\;\;\;\;\;\;\;\;\;  \hat{r}_{\frac{L}{2},+}^{ccw} &&=-\kappa \, \hat{b}_{in} + \tau \, \hat{r}_{\frac{L}{2},-}^{ccw}.
\end{alignat}

These first four equations are the familiar directional coupler equations for the counter clockwise circulating $\hat{a}$ and $\hat{b}$ modes. These next four equations are for the clockwise circulating $\hat{c}$ and $\hat{d}$ modes and are formally the same, except that they travel through the MRR in the opposite direction: 

% \begin{align}
%     \hat{c}_{out}=\tau \, \hat{c}_{in} + \kappa \, \hat{r}_{0}^{cw}, \\
%     \hat{r}_{L}^{cw}=-\kappa \, \hat{c}_{in} + \tau \, \hat{r}_{0}^{cw}, \\
%     \hat{d}_{out}= \tau \, \hat{d}_{in} + \kappa \, \hat{r}_{\frac{L}{2},+}^{cw}, \\
%     \hat{r}_{\frac{L}{2},-}^{cw}=-\kappa \, \hat{d}_{in} + \tau \, \hat{r}_{\frac{L}{2},+}^{cw}.    
% \end{align}
\begin{alignat}{3}
    \hat{c}_{out}&=\tau \, \hat{c}_{in} + \kappa \, \hat{r}_{0}^{cw}, \;\;\;\;\;\;\;\;\;\;\;\;\;\;\;\;\hat{r}_{L}^{cw}&&=-\kappa \, \hat{c}_{in} + \tau \, \hat{r}_{0}^{cw}, \nonumber \\
    \hat{d}_{out}&= \tau \, \hat{d}_{in} + \kappa \, \hat{r}_{\frac{L}{2},+}^{cw},\;\;\;\;\;\;\;\;\;\;\;\;\; \hat{r}_{\frac{L}{2},-}^{cw}&&=-\kappa \, \hat{d}_{in} + \tau \, \hat{r}_{\frac{L}{2},+}^{cw}. 
\end{alignat}

 The interior mode equations that traverse through the beam splitters and allow for backscattering are as follows:
% \begin{align}
%     \hat{r}_L^{ccw}=\sqrt{\gamma} \, e^{i \frac{\theta}{2}} \, \hat{r}_{\frac{L}{2},+}^{ccw} +\sqrt{1-\gamma} \, e^{i \frac{\theta}{2}} \, \hat{r}_{L}^{cw}, \\
%     \hat{r}_{\frac{L}{2},+}^{cw}=\sqrt{\gamma} \, e^{i \frac{\theta}{2}} \, \hat{r}_{L}^{cw} -\sqrt{1-\gamma} \, e^{i \frac{\theta}{2}} \, \hat{r}_{\frac{L}{2},+}^{ccw}, \\
%     \hat{r}_{\frac{L}{2},-}^{ccw}=\sqrt{\gamma} \, e^{i \frac{\theta}{2}} \, \hat{r}_0^{ccw} +\sqrt{1-\gamma} \, e^{i \frac{\theta}{2}} \, \hat{r}_{\frac{L}{2},-}^{cw}, \\
%     \hat{r}_{0}^{cw}=\sqrt{\gamma} \, e^{i \frac{\theta}{2}} \, \hat{r}_{\frac{L}{2},-}^{cw} -\sqrt{1-\gamma} \, e^{i \frac{\theta}{2}} \, \hat{r}_0^{ccw}. 
% \end{align}
\begin{alignat}{3}
    \hat{r}_L^{ccw}&=\sqrt{\gamma} \, e^{i \frac{\theta}{2}} \, \hat{r}_{\frac{L}{2},+}^{ccw} +\sqrt{1-\gamma} \, e^{i \frac{\theta}{2}} \, \hat{r}_{L}^{cw}, \;\;\;\;\;\;\;\;
    \hat{r}_{\frac{L}{2},+}^{cw}&&=\sqrt{\gamma} \, e^{i \frac{\theta}{2}} \, \hat{r}_{L}^{cw} -\sqrt{1-\gamma} \, e^{i \frac{\theta}{2}} \, \hat{r}_{\frac{L}{2},+}^{ccw}, \nonumber\\
    \hat{r}_{\frac{L}{2},-}^{ccw}&=\sqrt{\gamma} \, e^{i \frac{\theta}{2}} \, \hat{r}_0^{ccw} +\sqrt{1-\gamma} \, e^{i \frac{\theta}{2}} \, \hat{r}_{\frac{L}{2},-}^{cw}, \;\;\;\;\;\;\;\;
    \hat{r}_{0}^{cw}&&=\sqrt{\gamma} \, e^{i \frac{\theta}{2}} \, \hat{r}_{\frac{L}{2},-}^{cw} -\sqrt{1-\gamma} \, e^{i \frac{\theta}{2}} \, \hat{r}_0^{ccw}. 
\end{alignat}

These equations also include the phase propagation of the internal modes within the MRR between the designated modes.
By convention we have introduced minus signs when photons reflect from counterclockwise modes into clockwise modes or when photons cross from the waveguide into the MRR. Now it should be clear that $\gamma$ represents the actual transmission through the internal beam splitter. As we present results for various values of $\gamma$, keep in mind that a plot with $\gamma=0.95$ means that 95\% of the photons are transmitting through the internal beam splitter and 5\% of photons are backscattering into the counter-propagating modes.

The above set of coupled mode equations produces, in general, a $4\times 4$ unitary transfer matrix $\mathcal{S}$ with overall transition amplitudes, in terms of the annihilation operators, of all the output modes resulting from the input modes.

% \begin{equation}
%     \begin{pmatrix}
%     \hat{a}_{out} \\
%     \hat{b}_{out} \\
%     \hat{c}_{out} \\
%     \hat{d}_{out}
%     \end{pmatrix}
%     =
%     \begin{pmatrix}
%     A_1 & A_2 & A_3 & A_4 \\
%     B_1 & B_2 & B_3 & B_4 \\
%     C_1 & C_2 & C_3 & C_4 \\
%     D_1 & D_2 & D_3 & D_4 
%     \end{pmatrix}
%     \begin{pmatrix}
%     \hat{a}_{in} \\
%     \hat{b}_{in} \\
%     \hat{c}_{in} \\
%     \hat{d}_{in}
%     \end{pmatrix}
%     \label{generalmatrixeq}
% \end{equation}

% \begin{equation}
%     \begin{pmatrix}
%     \hat{a}_{out} \\
%     \hat{b}_{out} \\
%     \hat{c}_{out} \\
%     \hat{d}_{out}
%     \end{pmatrix}
%     =
%     \mathcal{S}
%     \begin{pmatrix}
%     \hat{a}_{in} \\
%     \hat{b}_{in} \\
%     \hat{c}_{in} \\
%     \hat{d}_{in}
%     \end{pmatrix}
%     \label{generalmatrixeqS}
% \end{equation}
\begin{equation}
    \begin{pmatrix}
    \hat{a}_{out} \\
    \hat{b}_{out} \\
    \hat{c}_{out} \\
    \hat{d}_{out}
    \end{pmatrix}
    =
    \begin{pmatrix}
    A_1 & A_2 & A_3 & A_4 \\
    B_1 & B_2 & B_3 & B_4 \\
    C_1 & C_2 & C_3 & C_4 \\
    D_1 & D_2 & D_3 & D_4 
    \end{pmatrix}
    \begin{pmatrix}
    \hat{a}_{in} \\
    \hat{b}_{in} \\
    \hat{c}_{in} \\
    \hat{d}_{in}
    \end{pmatrix}
    =
    \mathcal{S}
    \begin{pmatrix}
    \hat{a}_{in} \\
    \hat{b}_{in} \\
    \hat{c}_{in} \\
    \hat{d}_{in}
    \end{pmatrix}
    \label{generalmatrixeq}
\end{equation}

For propagation of quantum states through the MRR, 
we follow the work of Skaar, et al. (2004) \cite{Skaar:2004} and rearrange this matrix equation to obtain the transition amplitudes, in terms of the creation operators, for the input modes in terms of the output modes.

\begin{equation}
    \begin{pmatrix}
    \hat{a}_{in}^\dagger \\
    \hat{b}_{in}^\dagger \\
    \hat{c}_{in}^\dagger \\
    \hat{d}_{in}^\dagger
    \end{pmatrix}
    =
    \mathcal{S}^T
    \begin{pmatrix}
    \hat{a}_{out}^\dagger \\
    \hat{b}_{out}^\dagger \\
    \hat{c}_{out}^\dagger \\
    \hat{d}_{out}^\dagger
    \end{pmatrix}
    \label{generalmatrixeqST}
\end{equation}

In practice, we will only use a $2\times 2$ sub-matrix of this general $4\times 4$ matrix at any  time (see \App{Appendix:1}). In this work, we will always input photons on two different modes, and similarly detect on only two distinct output modes. We will first consider the case of inputting one photon each into $\hat{a}_{in}^\dagger$ and $\hat{b}_{in}^\dagger$ and looking for coincidence counts on $\hat{a}_{out}^\dagger$ and $\hat{b}_{out}^\dagger$. These are the original input and output modes used when there is no backscattering, and we will examine the effect on the crescent shape
\cite{Hach:2014,Alsing_Hach:2017a,Kaulfuss_thesis:2021,Kaulfuss:2023.Identical}
% FIX ARXIV CITATION ONCE WE %GET IT PUBLISHED
of the HOMM when we change the value of the backscattering parameter $\gamma$. 
We will focus on $\hat{a}_{out}^\dagger$ and $\hat{b}_{out}^\dagger$ outputs with $\hat{a}_{in}^\dagger$ and $\hat{b}_{in}^\dagger$ inputs (AB input/output), which is equivalent to (CD input/output). 
In this section, we will analyze the AB input/output results (note: CD input/output results are identical) since the HOM effect is present without backscattering, and we will investigate the effect backscattering has on the HOMM. It is possible to achieve the HOM effect for different input/output combinations, but it requires a high level of backscattering that would be unrealistic to expect without specific design and fabrication for intentional backscattering. We will discuss these other combinations of inputs and outputs and how the HOM effect can be achieved in \Sec{otherinout}. As a brief summary: in our current backscattering model the overall transfer matrix relating the inputs to the outputs is, in general, a $4\times 4$. For the case of both inputs and outputs on the AB modes,  we will focus on only the $2\times 2$ sub-matrix relating $\hat{a}_{in}^\dagger$ and $\hat{b}_{in}^\dagger$ with $\hat{a}_{out}^\dagger$ and $\hat{b}_{out}^\dagger$, involving the MRR transfer matrix elements $A_1$, $A_2$, $B_1$, and $B_2$.  
\begin{equation}
    \begin{pmatrix}
    \hat{a}_{in}^\dagger \\
    \hat{b}_{in}^\dagger
    \end{pmatrix}
    =
    \begin{pmatrix}
    A_1 & B_1 \\
    A_2 & B_2
    \end{pmatrix}
    \begin{pmatrix}
    \hat{a}_{out}^\dagger \\
    \hat{b}_{out}^\dagger
    \end{pmatrix}.
\end{equation}
Note that the above matrix is the transpose of the upper left quadrant of the full unitary transfer matrix $\mathcal{S}$ in \Eq{generalmatrixeq}.
The transfer matrix describes the relationship between the inputs and outputs and the matrix elements are calculated by solving the system of boundary condition equations outlined above. The probability of detecting a particular output state with a given input state will be denoted as $P_{\{\text{out}|\text{in}\}}$. The probability of a coincident output (one photon in each output mode) with a single photon input on each input mode will therefore be written as: $P_{\{1,1|1,1\}}$ and is the permanent of the $2\times 2$ sub-matrix squared. 

\begin{equation}
    P_{\{1,1|1,1\}}=|A_1\,B_2+A_2\,B_1|^2 \label{PABAB}
\end{equation}

The HOM effect occurs when $P_{\{1,1|1,1\}}=0$. In the case of MRRs, this equation yields a 1D manifold of solutions which we call the HOMM.

\section{Single MRR Results} \label{SingleMRR}
\subsection{Preliminaries}
\begin{figure}[H]
    \centering
    \includegraphics[width=0.49\linewidth,keepaspectratio]{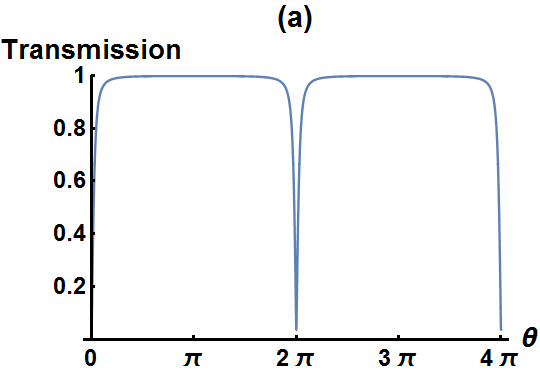}
    \includegraphics[width=0.49\linewidth,keepaspectratio]{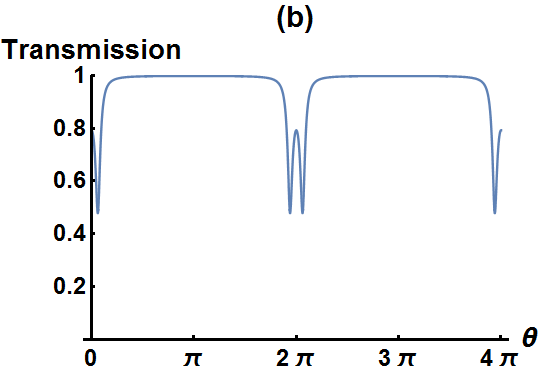}
    \caption{Transmission plots of $\hat{a}_{out}^\dagger$ with 1 photon input on $\hat{a}_{in}^\dagger$ and $\tau=0.95$ for (a) $\gamma=1$ (no backscattering) and (b) $\gamma=0.99$ (1\% backscattering).}
    \label{transmission}
\end{figure}

We begin by checking our model against MRR and backscattering results discussed in the literature. A very commonly reported effect of backscattering in MRRs is resonance splitting \cite{BackscatterReview:Li:2016,McCutcheon:2021,Hance:2021}. A simple test of our model that demonstrates that it does capture the resonance splitting phenomena is shown in the normalized transmission plots in \Fig{transmission}. These  plots examine the output on $\hat{a}_{out}^\dagger$ for a single photon input on $\hat{a}_{in}^\dagger$ when $\tau=0.95$ for (a) $\gamma=1$ (no backscattering) and (b) $\gamma=0.99$ (1\% backscattering).

It is clear to see from \Fig{transmission} that the MRR is resonant at $\theta=0, 2\pi, 4\pi\ldots$. This backscattering model produces symmetric resonance splitting, which is expected of internal ring backscattering due to sidewall roughness \cite{BackscatterReview:Li:2016,aym:splitting:note}. 
However, for the purposes of investigating the effect of backscattering on the HOMM we instead focus on the $\theta$ range where the MRR is off resonance (around $\theta=\pi$) \cite{Kaulfuss_thesis:2021,Kaulfuss:2023.Identical}. The transmission plot is unaffected by backscattering at $\theta$ values where the MRR is off resonance. 

Since photons can escape the system to alternate output modes, we must be careful about how we define the HOMM. Note that the condition $P_{\{1,1|1,1\}}=0$  is trivial if there are no photons exiting the relevant detector output ports and is not indicative of the HOM effect. It is also important to keep in mind that even when an HOMM is present it may not occur with 100\% probability.  Therefore, we must also ensure that for a given set of parameters there are photons being detected at the chosen output modes with nonzero probability (blue curve in figures below) in addition to the HOMM condition $P_{\{1,1|1,1\}}=0$.

\begin{figure}[H]
    \centering
    \includegraphics[width=0.5\linewidth,keepaspectratio]{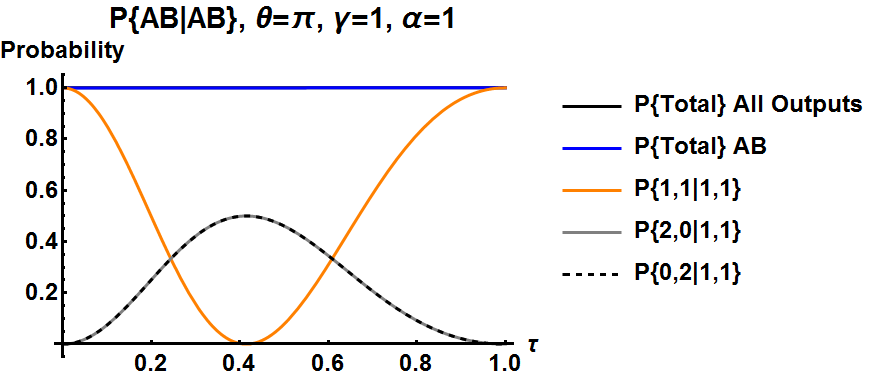}
    \caption{Probability plot for a single MRR with AB inputs and AB outputs and no backscattering $(\gamma=1)$.}
    \label{exampleprobplot}
\end{figure}

\begin{figure}[H]
    \centering
    \includegraphics[width=0.35\linewidth,keepaspectratio]{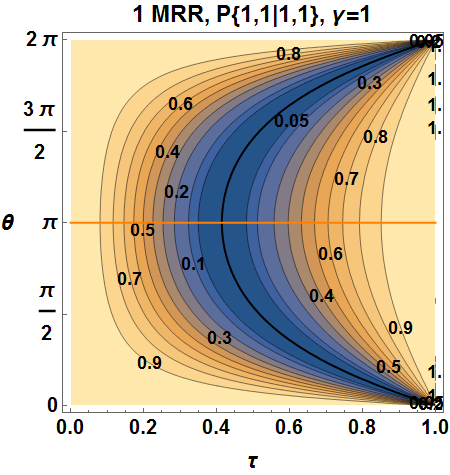}
    \caption{Contour plot of $P_{\{1,1|1,1\}}$ on AB output/inputs with $\gamma=1$. Black line is the exact HOMM curve and orange line represents the slice across $P_{\{1,1|1,1\}}$ at $\theta=\pi$, coinciding with the orange curve in each probability plot.}
    \label{examplewithlines}
\end{figure}

For the above reasons we have created probability plots (see \Fig{exampleprobplot}), which are a slice across $\theta=\pi$ (off resonance), to compare the relevant probabilities and be used in combination with the contour plots (see \Fig{examplewithlines}). 

In \Fig{exampleprobplot} the black solid line is the overall probability considering all possible outputs across all four output modes, and should therefore always be unity. The solid blue line (which overlaps the black line here) is the overall cumulative probability of detecting all possible outputs on the two modes considered (here, AB). The solid orange line is the probability of detecting a 
$\{1,1\}$ output on the two detected modes, and the solid gray and dashed black lines are the probabilities of detecting $\{2,0\}$ and $\{0,2\}$ outputs on the two detected modes. The most important characteristic to look for to identify the HOMM is a dip to zero in the orange curve, while the gray and dashed black curves (and subsequently the blue curve) are non-zero, indicating that there are still photons exiting the detected modes in the form of $\{2,0\}$ and $\{0,2\}$ states. 

The contour plots in this section are for the probability $P_{\{1,1|1,1\}}$ of detecting a 1,1 output coincidence measurement in the AB outputs, with a 1,1 input in the AB inputs. The abscissa $\tau$ is the linear coupling between the waveguide and the MRR, and the ordinate $\theta$ is the round-trip phase shift of the MRR. These are the two variables that differentiate the rings when we assume that the directional couplers are identical. In the contour plots, the contours are in probability, namely $P_{\{1,1|1,1\}}$. The darkest blue region is where the coincidence probability is between 0 and 0.05, which is the experimentally relevant region to find the HOM effect. Based on previous work, the optimal area to operate this device to obtain the HOM effect is in the center of the dark blue region, at $\theta=\pi$ and $\tau=\tau_{c}=0.4142$ for a single MRR \cite{Kaulfuss_thesis:2021,Kaulfuss:2023.Identical}. In the following example, the solid black curve on the contour plot is the exact HOMM for a lossless single MRR (the exact analytic solution for the black curve is given as Eq.(4) in Kaulfuss, et al. (2022) \cite{Kaulfuss:2023.Identical}) and the orange line on the contour plot across $\theta=\pi$ (see \Fig{examplewithlines}) coincides with the orange curves in the probability plot (see \Fig{exampleprobplot}), which shows the relationship between the two plots.

In addition to backscattering, we also model the effects of radiative loss along the MRR.
\Fig{probplotshowingloss} shows an example probability plot of $P_{\{1,1|1,1\}}$ on AB output/input ports with $\gamma=0.9$ for no loss ($\alpha=1$) and 10\% internal radiative loss ($\alpha=0.9)$:
\begin{figure}[H]
    \centering
    \includegraphics[width=0.31\linewidth,keepaspectratio]{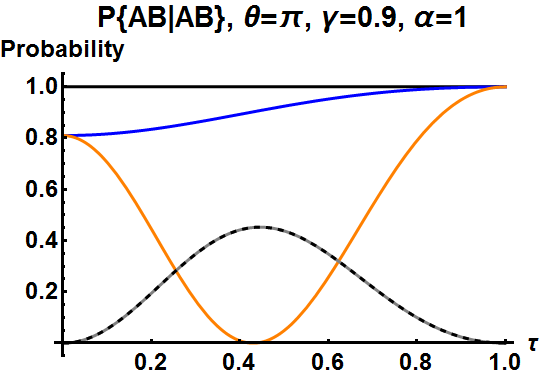}
    \includegraphics[width=0.5\linewidth,keepaspectratio]{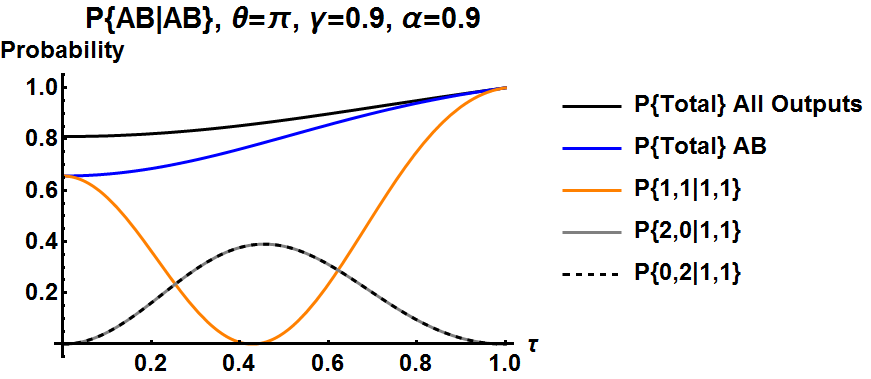}
    \caption{Probability plots of $P_{\{1,1|1,1\}}$ on AB output/inputs with $\gamma=0.9$ and $\alpha=1$ (left) and $\alpha=0.9$ (right).}
    \label{probplotshowingloss}
\end{figure}
\noindent Here we have added phenomenological round-trip radiative loss $0\le\alpha\le 1$, to each exponential term (i.e. $e^{i \theta} \xrightarrow{}  e^{i (\theta + i\,\tilde{\theta})}\equiv \alpha\,e^{i \theta} $) as described in Yariv (2000)\cite{Yariv:2000}.  As shown in \Fig{probplotshowingloss}, loss simply causes all the curves to shift downwards as some probability goes undetected (modeling scattering into undetected internal modes within the MRR). This loss is weighted towards lower $\tau$ values because it represents propagation loss in the MRR and as $\tau$ increases more photons are transmitting through the waveguide without coupling into the MRR. This does not cause any difference to the darkest blue (HOM effect) regions in the contour plots, but it should be kept in mind that this would cause an experimenter to have to detect for a longer time in order to observe the HOM effect due to the reduction of the overall probability of detecting photons on the measured outputs (blue curve). A full quantum mechanical treatment of loss on the HOMM for MRRs was presented in Alsing, et al. \cite{Alsing_Hach:2017a} and showed that the phenomenological loss model of Yariv is adequate for simulating loss. 
As such, for the rest of this investigation into the effect of backscattering on HOMM in MRRs we will consider only the lossless situation here, $\alpha=1$.

%(for plots with $\alpha<1$, see \App{Appendix:2}).  

\subsection{Contour plots of HOMM probability for a single MRR}

We will now show probability and contour plots for a single MRR across a range of $\gamma$ values: 1, 0.99, 0.95, and 0.9. A realistic level of backscattering is around or less than 1\%\cite{Ballesteros:2011,BackscatterReview:Li:2016,McCutcheon:2021}, but we have also included 5\% and 10\% backscattering as extreme examples to show the robustness of the HOMM in MRRs.   

%%%%%%%%%%%%%%%%%%%%%%%%%%%%%%%%%%%%%%%%%%%%%%%%%%%%%%%%%%%
% COMMENTING OUT THESE FIGURES AND COMBINING INTO ONE WITH CPLOTS
%\begin{figure}[H]
%    \centering
%    \begin{tabular}{cc}
%    \includegraphics[width=0.39\linewidth,keepaspectratio]{probg=1a=1nolabel.png} & 
%    \includegraphics[width=0.61\linewidth,keepaspectratio]{probg=0.99a=1.png}  \\
%    \includegraphics[width=0.39\linewidth,keepaspectratio]{probg=0.95a=1nolabel.png} & \includegraphics[width=0.61\linewidth,keepaspectratio]{probg=0.9a=1.png}
%    \end{tabular}
%    \caption{Single MRR Probability Plots $\gamma=1,0.99,0.95,0.9$}
%    \label{1MRRprobplots}
%\end{figure}

%%%% COMMENTING OUT AND COMBINING WITH PROB PLOTS

%\begin{figure}[H]
%    \centering
%    \begin{tabular}{cc}
%    \includegraphics[width=0.45\linewidth,keepaspectratio]{contourg=1.png} & 
%    \includegraphics[width=0.45\linewidth,keepaspectratio]{contourg=0.99.png}  \\
%    \includegraphics[width=0.45\linewidth,keepaspectratio]{contourg=0.95.png} & \includegraphics[width=0.45\linewidth,keepaspectratio]{contourg=0.9.png}
%    \end{tabular}
%    \caption{Single MRR Contour with backscattering $\gamma=1,0.99,0.95,0.9$}
%    \label{backreflectiongamma1}
%\end{figure}

%0.3,0.5

\begin{figure}[H]
    \centering
    \begin{tabular}{m{1.9in}m{3.0in}}
    \includegraphics[width=\linewidth,keepaspectratio]{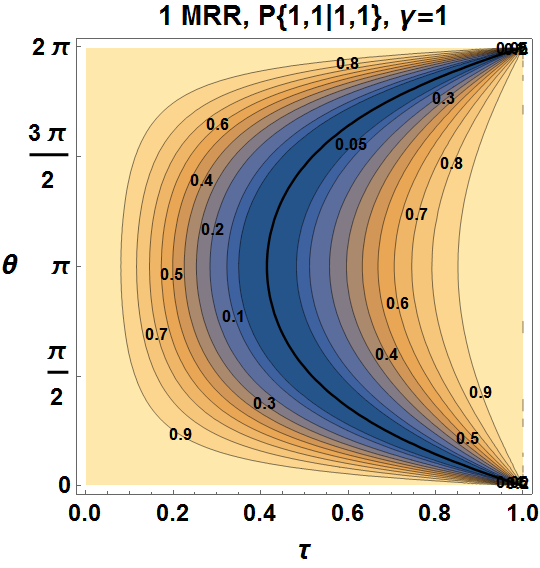} & 
    \includegraphics[width=1.2\linewidth,keepaspectratio]{probg=1a=1.png} \\
    \includegraphics[width=\linewidth,keepaspectratio]{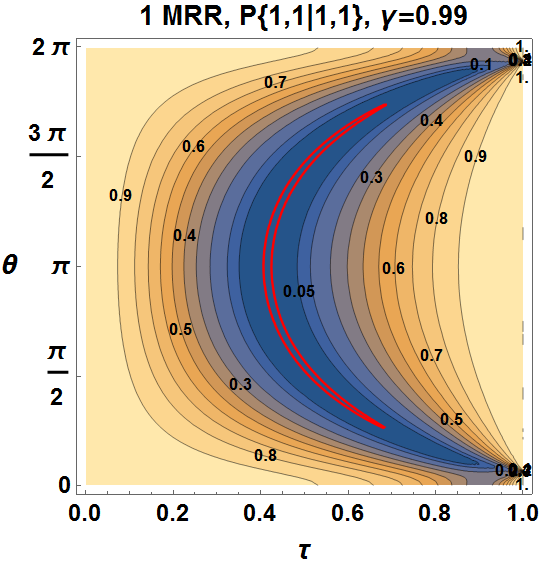} & \includegraphics[width=1.2\linewidth,keepaspectratio]{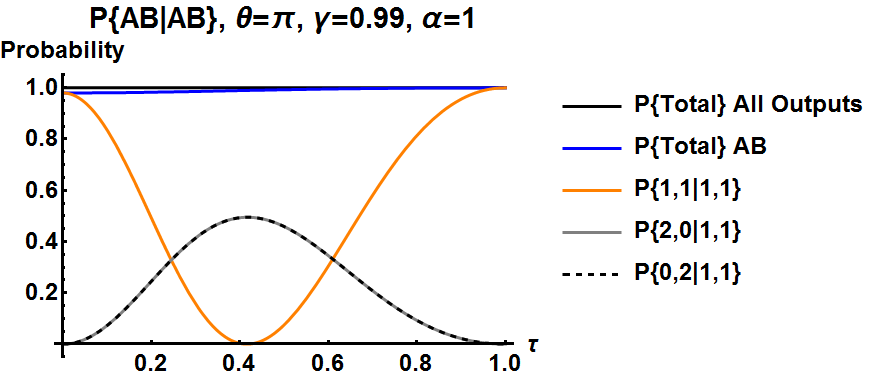} \\
    \includegraphics[width=\linewidth,keepaspectratio]{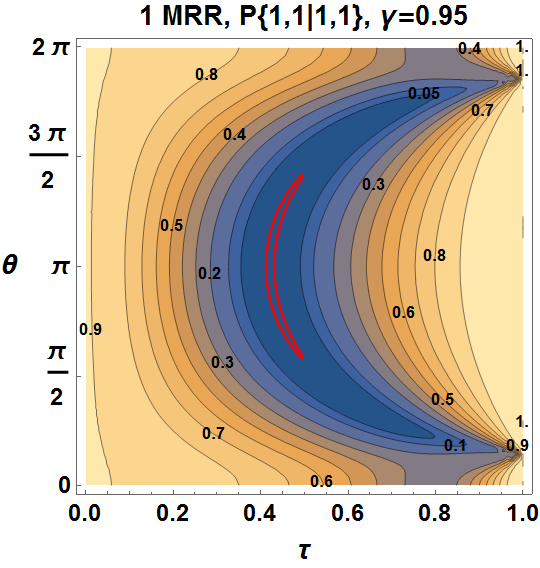} & \includegraphics[width=1.2\linewidth,keepaspectratio]{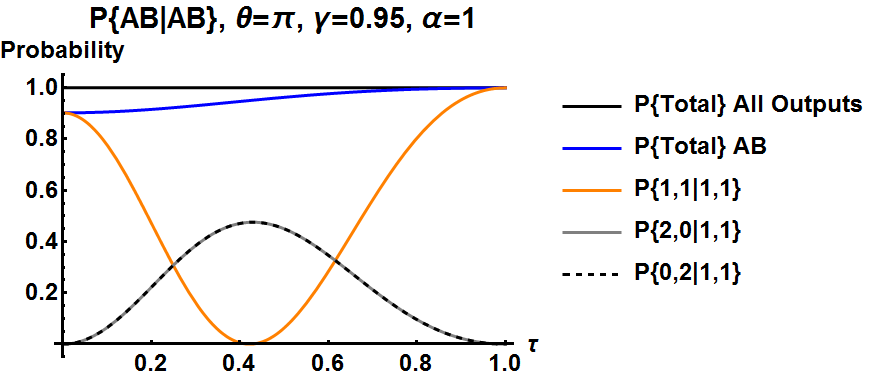} \\
    \includegraphics[width=\linewidth,keepaspectratio]{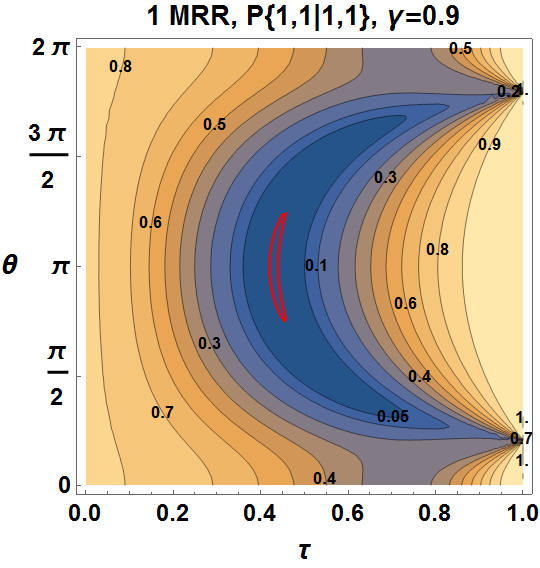} & \includegraphics[width=1.2\linewidth,keepaspectratio]{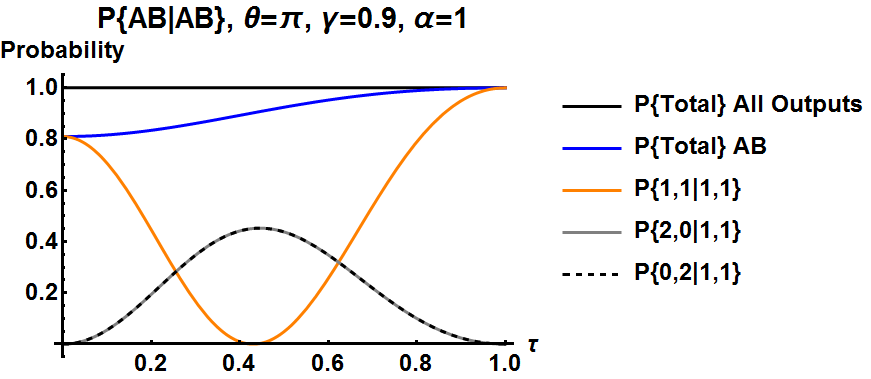}
    \end{tabular}
    \caption{Contour (left) and probability (right) plots for a single MRR with $\gamma= (1,0.99,0.95,0.9)$. These plots show the effect of backscattering on the HOMM for a single MRR with AB inputs/outputs. The red contour is a probability of $10^{-3}$.}
    \label{comboplots1MRR}
\end{figure}
\clearpage
\newpage

Notice in \Fig{comboplots1MRR}, that the HOMM dip of $P_{\{1,1|1,1\}}$ (orange curve, right column) is 
maintained and the probability is only qualitatively reduced by backscattering at very low $\tau$ values, where most (or all) of the photons are coupled into the MRR at the directional couplers. Also, the overall detection probability in the AB outputs (blue curve) is reduced as $\gamma$ is increased, representing more photons being `lost' to output modes we are not detecting on. Practically, this means that the HOM effect would still be seen detecting on the AB outputs, one would simply have to wait longer to see a successful event.  

These contour plots represent $P_{\{1,1|1,1\}}$ across the entire range of $\tau$ and $\theta$ values, and the orange curve in the probability plots represents a slice across $\theta=\pi$ on these contour plots as shown in \Fig{examplewithlines}. The contour plot for $\gamma=1$ is identical to our original single MRR identical contour plot (with no backscattering) \cite{Kaulfuss_thesis:2021}. As it should, because as the backscattering approaches $\gamma=1$, we should recover the ideal single MRR results with no backscattering. The $\gamma=1$ plot in (\Fig{comboplots1MRR}) has a single black curve representing the exact analytic HOMM solution for a single MRR \cite{Kaulfuss:2023.Identical,Kaulfuss_thesis:2021}. However, once backscattering is introduced to the system we can no longer achieve an exact analytic solution for the HOMM, so we plot a red contour at a very low probability, $P_{\{1,1|1,1\}}=10^{-3}$. The HOMM (where the probability is zero exactly) is a one dimensional line bounded by the red contour in each contour plot.

\begin{figure}[H]
    \centering
    \begin{tabular}{m{2in}m{3.2in}}
    \includegraphics[width=\linewidth,keepaspectratio]{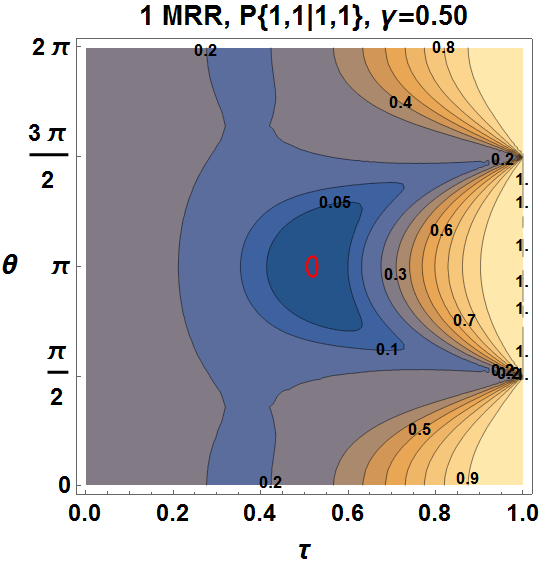} &
    \includegraphics[width=1.2\linewidth,keepaspectratio]{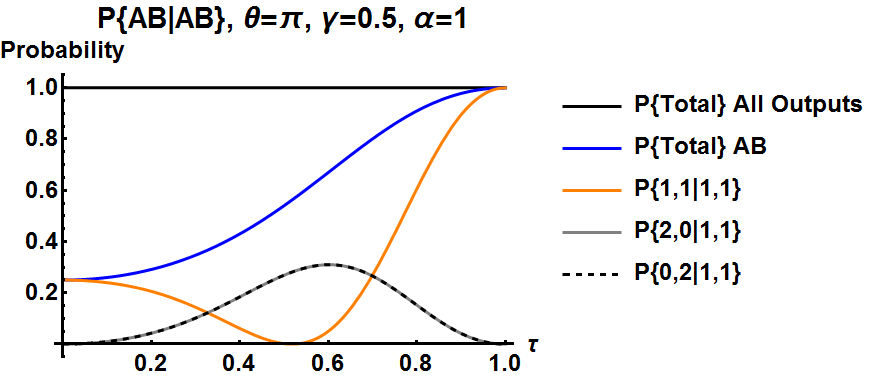}
    \end{tabular}
    \caption{Single MRR Contour and probability plots with backscattering $\gamma=0.5$. The red contour is a probability of $10^{-3}$.}
    \label{1mrrgamma0.5}
\end{figure}

A $\gamma$ value of 0.9 (10\% backscattering) is already beyond what we would expect to see in experiment. In \Fig{1mrrgamma0.5} we show an example with extreme backscattering of $\gamma=0.5$, which means that half of the photons are backscattering at each internal beam splitter. The fact that the HOMM experimental region persists, even in this extreme case, is clear evidence of the robustness of the HOMM in MRRs against backscattering.
\noindent The contour plots of \Fig{comboplots1MRR} and \Fig{1mrrgamma0.5} show that as the backscattering increases the HOMM shrinks from a 1D manifold (curve) at $\gamma=1$ (no backscattering), to an effective 0D manifold (point) as $\gamma\to 0$ ($100\%$ backscattering).

Notice that even as backscattering is increased to extremely high values, the optimal operating region of the crescent ($\theta=\pi$ and $\tau=\tau_{c}$)\cite{Kaulfuss:2023.Identical} is preserved, just shifted to slightly larger $\tau$ values. The only regions that are lost are at the edges of the crescent, which are also at the edges of the $\theta$ range and high $\tau$ values, where we wouldn't expect to be experimentally operating regardless \cite{Kaulfuss:2023.Identical}. Note that the HOMM curve is shifted to slightly higher $\tau$ values for high levels of backscattering, but for the experimentally relevant levels of backscattering shown in (\Fig{comboplots1MRR}) the shift is very small. 

\section{Alternate Input/Output Combinations}\label{otherinout}
We now investigate some other combinations of inputs and outputs for a single MRR (\Fig{back reflection MRR diagram}) using our general $4\times 4$ matrix and see if the HOM effect can still be achieved for combinations other than $\text{AB}_{\text{out}}\text{AB}_{\text{in}}$ or $\text{CD}_{\text{out}}\text{CD}_{\text{in}}$. Particularly, we want to investigate if the HOM interference effect can occur when detecting on backscattered modes. Remember that the notation for probability is the probability to detect an output with a given input is denoted as
$P_{\{\text{out}|\text{in}\}}$.  

\subsection[CDoutABin] {$\text{CD}_{\text{out}}\text{AB}_{\text{in}}$: ccw input, cw output} \label{CDoutABin}

If we input on the traditional AB (ccw) input , but detect on the new backscattered CD (cw) outputs we  
\begin{figure}[H]
    \centering
    \begin{tabular}{cc}
    \includegraphics[width=0.45\linewidth,keepaspectratio]{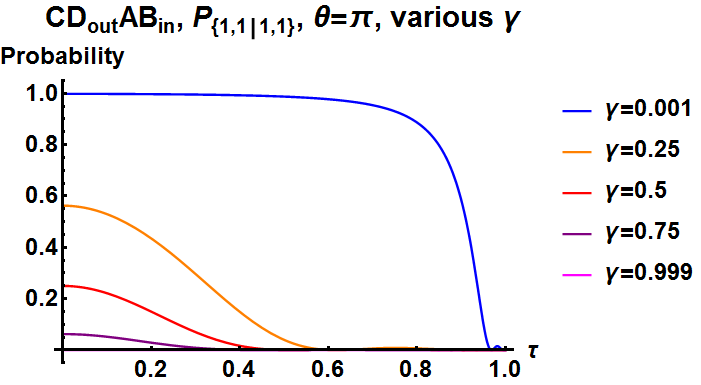} &
    \includegraphics[width=0.45\linewidth,keepaspectratio]{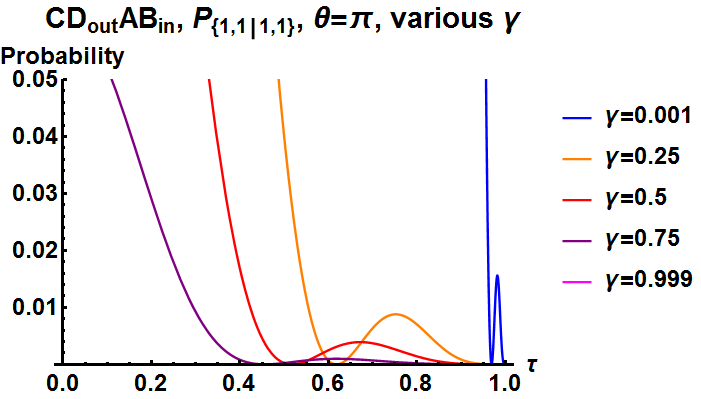}
    \end{tabular}
    \caption{Entire range probability plot (left) and zoomed probability plot between 0 and 0.05 (right) of $P_{\{1,1|1,1\}}$ with AB input and CD output for various $\gamma$ values (0.001, 0.25, 0.5, 0.75, 0.999).}
    \label{ABinCDout}
\end{figure}
\noindent find that the HOM effect can still be detected, but high levels of backscattering are required. Keep in mind that we expect normal levels of backscattering to be around or less than 1\%\cite{Ballesteros:2011,BackscatterReview:Li:2016,McCutcheon:2021}. The amount of backscattering required in this section to detect the HOM effect for different input/output combinations is far beyond what we would reasonably expect to see in a normally fabricated MRR, as shown in \Fig{ABinCDout}.
Notice that dips to zero probability exist in \Fig{ABinCDout} for various $\gamma$, but high levels of backscattering are required for this interference effect to be detectable. 

In \Fig{ABinCDout} notice that there is no probability curve for $\gamma=0.999$ because as $\gamma\xrightarrow{}1$ there is no backscattering in the MRR and all the photons transmit through the interior beam splitters. Therefore, in our model when $\gamma\xrightarrow{}1$ there is no way for photons input on the AB modes to backscatter and output to the C or D outputs. Conversely, when $\gamma\xrightarrow{}0$ the interior beam splitters act as mirrors and all photons are backreflected into the C and D outputs, so the probability is 1 for all $\tau$ values, until a sharp drop off as $\tau\xrightarrow{}1$, because at $\tau\equiv 1$ 
the photons input at the A and B modes transmit straight through the waveguide without coupling into the MRR and have no chance to interact with the backscattering beam splitters. 

\begin{figure}[H]
    \centering
    \begin{tabular}{cc}
    \includegraphics[width=0.45\linewidth,keepaspectratio]{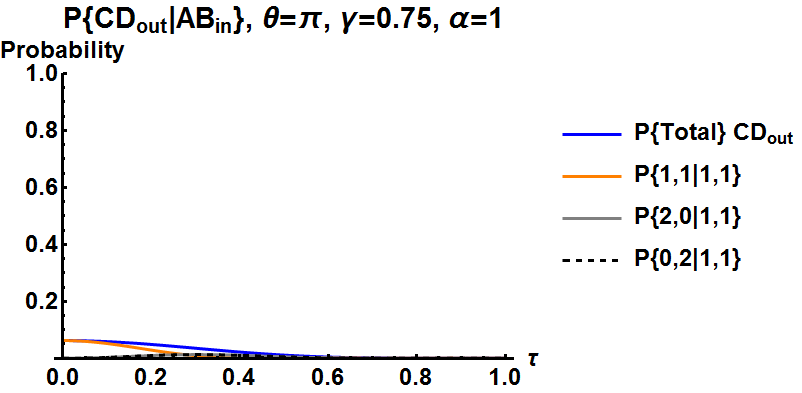} &
    \includegraphics[width=0.45\linewidth,keepaspectratio]{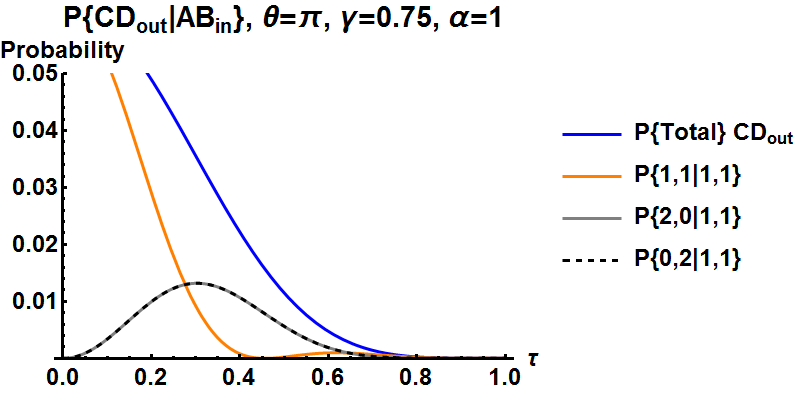} \\
    \includegraphics[width=0.45\linewidth,keepaspectratio]{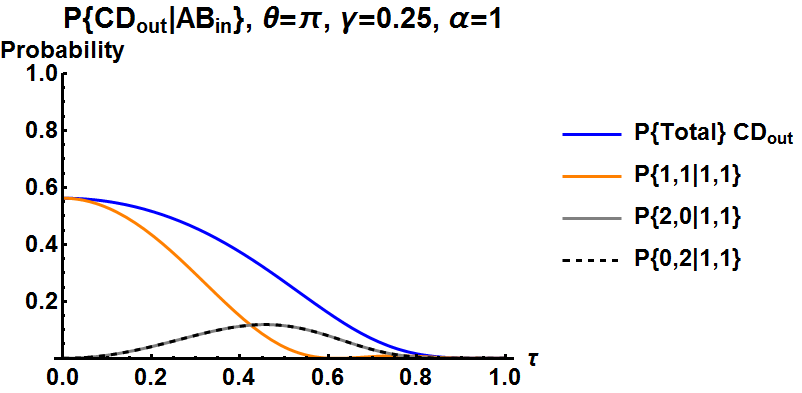} &
    \includegraphics[width=0.45\linewidth,keepaspectratio]{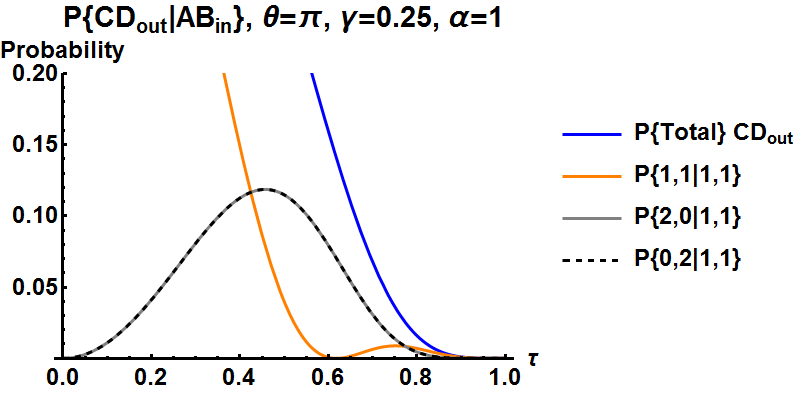}
    \end{tabular}
    \caption{Full probability plots of AB input and CD output for $\gamma=0.75$ (upper) and $\gamma=0.25$ (lower). The entire range is plotted on the left and a zoomed view to see where $P_{\{1,1|1,1\}}=0$ is plotted on the right (notice the range of probability values on the vertical axis).}
    \label{ABinCDoutfull}
\end{figure}

The curves in \Fig{ABinCDout} (right) exhibit dips to zero probability for $P_{\{1,1|1,1\}}$, i.e. a $\tau$ value for which the HOM effect will be obtained when $\theta=\pi$. However, even though the HOM effect will always occur at that set of $(\tau,\theta,\gamma)$ as long as photons exit the C and D outputs, we must construct the full probability plots to obtain an understanding of how often this would actually occur, since it is possible for photons to output at modes we are not detecting on. \Fig{ABinCDoutfull} shows the full probability plots corresponding to \Fig{ABinCDout},including all output states for $\gamma=0.75$ and $\gamma=0.25$. 

The full probability plots of \Fig{ABinCDoutfull} give a much clearer picture of how often one should expect to see the HOM effect on the backscattered modes. The blue curve is the probability that when you input photons on A and B modes that you will detect photons on C and D modes in any of the three possible output states. The orange curve is the $P_{\{1,1|1,1\}}$ plotted above in \Fig{ABinCDout} and the solid gray and dashed black lines are the \{2,0\} and \{0,2\} output states on the CD output. The most important information obtained from these plots is the probability value of the blue curve at the $\tau$ value where the orange curve has a dip in probability to zero. For example in the upper right plot of \Fig{ABinCDoutfull}, for $\gamma=0.75$ (25\% backscattering) the blue curve has a probability value of about 0.025 at the $\tau$ value where the orange curve dips to 0 probability. This means that if the photons output at the C and D modes for these parameter values $(\tau,\theta,\gamma)$ you will achieve the HOM effect with certainty, but this only occurs 2.5\% of the time. For an extreme level of 75\% backscattering (bottom right plot of \Fig{ABinCDoutfull}), the HOM effect could be detected with a probability of about 15\%. So for realistic levels of backscattering, less than 1\% ($\gamma\geq0.99$), the chance of detecting the HOM effect on the backscattered modes is effectively zero. In order to achieve the HOM effect on the backscattered modes with a noticeable probability an MRR would have to be designed and fabricated with greater than 50\% backscattering.

\subsection[ADoutADin] {$\text{AD}_{\text{out}}\text{AD}_{\text{in}}$: ccw and cw in, ccw and cw out}\label{ADoutADin}

\begin{figure}[H]
    \centering
    \begin{tabular}{cc}
    \includegraphics[width=0.45\linewidth,keepaspectratio]{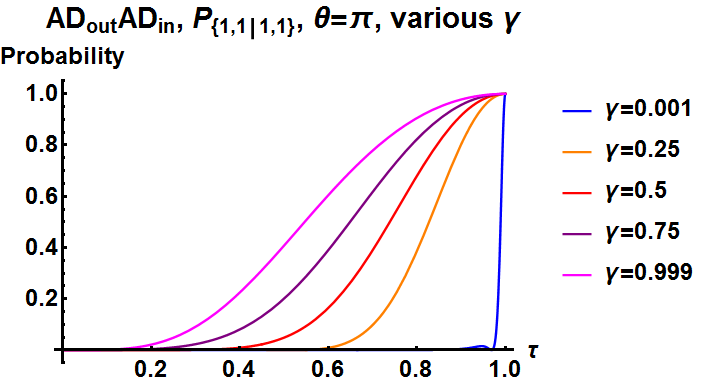} &
    \includegraphics[width=0.45\linewidth,keepaspectratio]{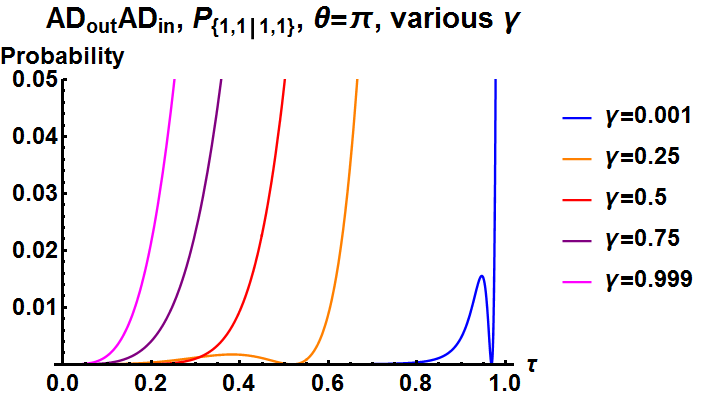}
    \end{tabular}
    \caption{Entire range probability plot (left) and zoomed probability plot between 0 and 0.05 (right) of $P_{\{1,1|1,1\}}$ with AD input and AD output for various $\gamma$ values (0.001, 0.25, 0.5, 0.75, 0.999).}
    \label{ADinADout}
\end{figure}

Similarly, for a different input/output combination, AD inputs and AD outputs,  the HOM effect can only be achieved for even higher levels of backscattering. Whereas in \Fig{ABinCDout} for $\text{CD}_{\text{out}}\text{AB}_{\text{in}}$ we see dips in $P_{\{1,1|1,1\}}$ to zero beginning at $\gamma=0.75$, for $\text{AD}_{\text{out}}\text{AD}_{\text{in}}$ (\Fig{ADinADout}) the first dip to zero occurs for $\gamma=0.25$. 

\begin{figure}[H]
    \centering
    \includegraphics[width=0.45\linewidth,keepaspectratio]{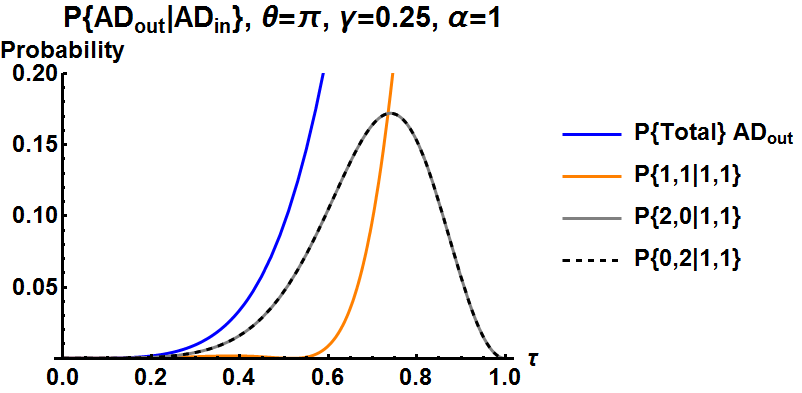} 
    \caption{Full probability plot of AD inputs and AD outputs for $\gamma=0.25$. Zoomed to show relevant probabilities for the HOM effect.}
    \label{probADinADoutg0.25}
\end{figure}

For $\text{AD}_{\text{out}}\text{AD}_{\text{in}}$ 75\% backscattering is required to obtain the HOM effect, which is even harder to achieve than $\text{CD}_{\text{out}}\text{AB}_{\text{in}}$. This is because the HOM effect requires the complete destructive interference of the \{1,1\} state. In the case of $\text{CD}_{\text{out}}\text{AB}_{\text{in}}$ there is no way for photons to reach the outputs without the presence of backscattering. However, for $\text{AD}_{\text{out}}\text{AD}_{\text{in}}$ photons can reach the outputs without backscattering via direct transmission. This difference manifests in higher $P_{\{1,1|1,1\}}$ values for $\text{AD}_{\text{out}}\text{AD}_{\text{in}}$, thus requiring more backscattering in order for sufficient destructive interference to occur and result in the HOM effect.

This concludes our investigation of alternate input/output combinations. The conclusion is that it is possible to achieve the HOM effect for different combinations of forward propagating and backscattering modes, but for all combinations the level of backscattering required is far beyond current physical devices. Even for the extreme $\gamma=0.9$ (10\% backscattering) we propose in Section III, these alternate input/output combinations would not achieve the HOM effect with any detectable probability. MRR based devices would have to specifically be designed and manufactured with unrealistic levels of backscattering to see the HOM effect on these alternate modes.

%\subsection{ACinACout}

%\begin{figure}[H]
%    \centering
%    \includegraphics[width=0.45\linewidth,keepaspectratio]{ACinACoutoverall.png} 
%    \caption{ACinACout for a range of gamma values}
%    \label{ACinACout}
%\end{figure}

%If we include ACinACout its just to illustrate I guess the idea that if you input completely on the bottom or the top and turn the backscattering up you create basically a half ring cavity and the photons cannot escape, you also get new resonances on this new cavity at odd integer values of pi.

\section{Contour Plots for a (parallel) linear chain of two identical MRRs} \label{Parallel}
\begin{figure}[H]
    \centering
    \includegraphics[width=0.5\linewidth,keepaspectratio]{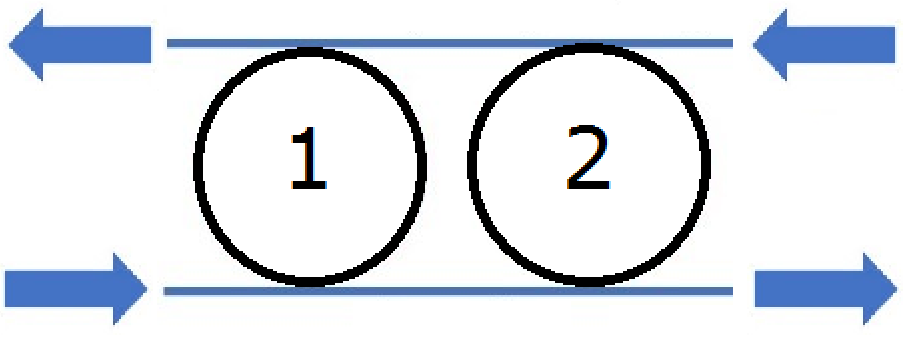}
    \caption{Diagram for 2 MRRs connected in parallel.}
    \label{2mrrparalleldiagram}
\end{figure}

\begin{figure}[H]
    \centering
    \begin{tabular}{cc}
    \includegraphics[width=0.35\linewidth,keepaspectratio]{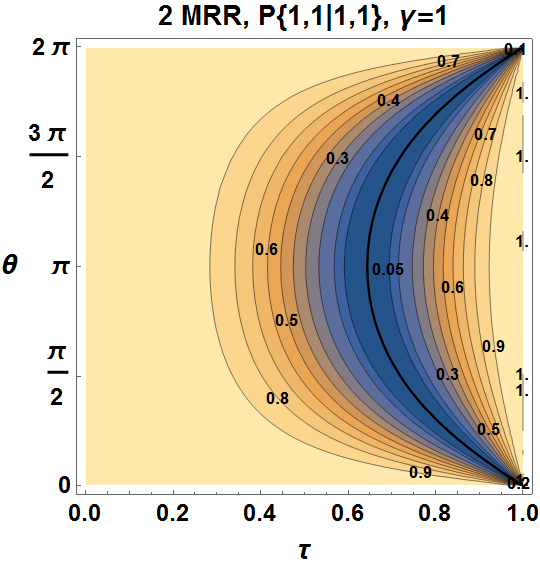} &
    \includegraphics[width=0.35\linewidth,keepaspectratio]{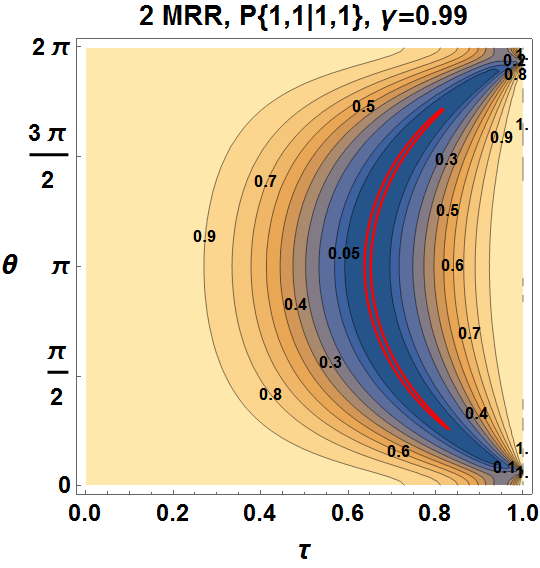} \\
    \includegraphics[width=0.35\linewidth,keepaspectratio]{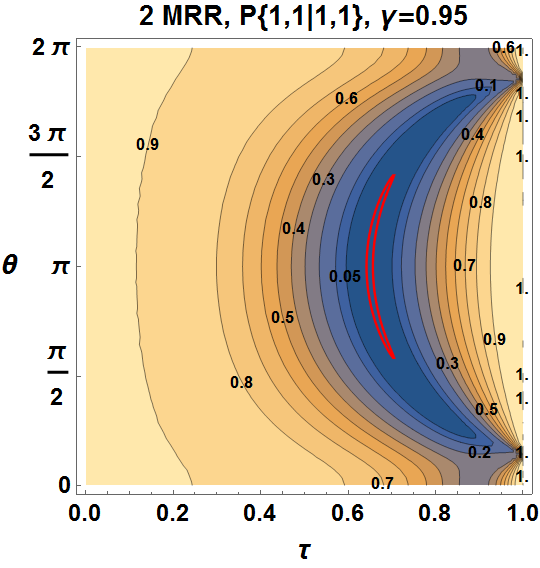} &
    \includegraphics[width=0.35\linewidth,keepaspectratio]{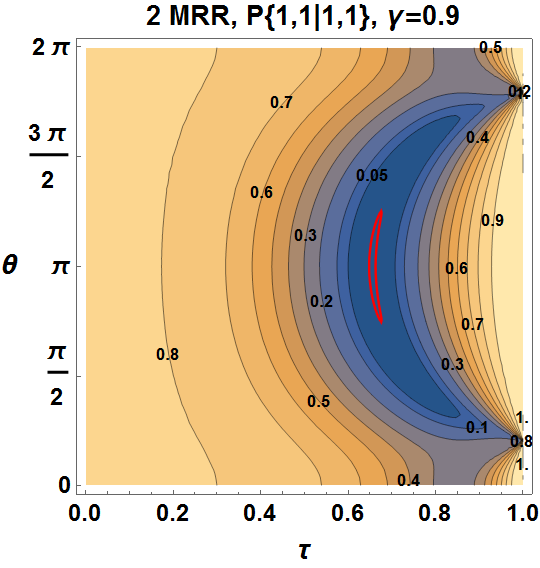}
    \end{tabular}
    \caption{2 MRR Parallel Contour with backscattering $\gamma=1,0.99,0.95,0.9$ (red contour is a probability of $10^{-3}$). The HOMM is a 1D curve bounded by the red contour plot.}
    \label{2mrrcontours}
\end{figure}

We will now investigate the effect of backscattering on a linear chain of identical MRRs in parallel\footnote{Note that the circuit shown in \Fig{2mrrparalleldiagram} was previously referred to by the authors of this paper as connected in ``series'' \cite{Kaulfuss:2023.Identical,Kaulfuss_thesis:2021}, but we are altering our terminology to match the literature and will now refer to circuits like these as connected in ``parallel'' \cite{Rabus:2007}.}
as shown in \Fig{2mrrparalleldiagram} to investigate if the noise due to backscattering is compounded causing the HOMM to deteriorate faster, or if the HOMM continues to persist against backscattering even in linear chains. For 2 MRRs in parallel we obtain the contour plots as shown in \Fig{2mrrcontours}.

Similar to the contour plots for the single MRR in \Fig{comboplots1MRR}, for two MRRs in parallel the edges of the contour plot of \Fig{2mrrcontours} are degraded, but the optimal operating area of the crescent shapes HOMM is preserved even for unrealistically high levels of backscattering. Once again, we include an extreme example with $\gamma=0.5$ for 2 identical MRRs in parallel. The HOMM persists even in a linear chain of MRRs in parallel, again showing a similar contraction of the 1D HOMM to a 0D HOMM for high levels of backscattering as $\gamma\to0$.
\begin{figure}[H]
    \centering
    \includegraphics[width=0.5\linewidth,keepaspectratio]{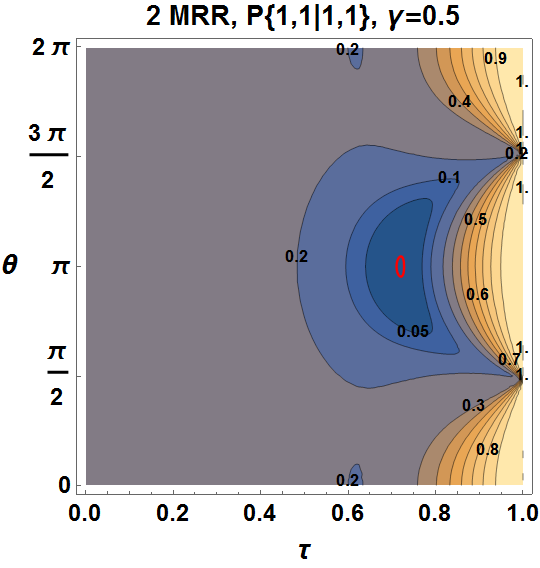}
    \caption{2 MRR Contour with backscattering $\gamma=0.5, \alpha=1$.}
    \label{2mrrgamma0.5}
\end{figure}
This trend in continues for 3 MRRs in parallel (not shown), maintaining the optimal operating region of the crescent but losing the edges as backscattering is increased. For higher numbers of MRRs in parallel the HOMM curves and subsequent contour crescents are shifted to higher values of $\tau$ \cite{Kaulfuss_thesis:2021}. A table of `figures of merit' of the HOMMs, including $\tau$ values, for up to 10 MRRs in parallel has been calculated using the exact analytic solution \cite{Kaulfuss:2023.Identical}. Therefore, we conclude that the HOMM is extremely robust against backscattering even in a linear chain of multiple backscattering MRRs as long as the MRRs are operated `off-resonance' $(\theta=\pi)$. 

\section{Results for a chain of two Non-Identical MRRs} \label{Nonidentical}

%%%%%%% OK I'm now asking myself why anyone cares that we remove the spike structure if its in an area that we already are saying not to operate in, yea it makes it "look" more like series of identical MRRs, but we are already only operating in the optimal region of tau=taumin and theta=pi, so why do we care whether the spike structures are there or not. 
%%%%%%%%%%%%%%%%%%%%%%%%%%%%%%%%%%%%%%%%%%%%%%%%%%%%%%%%%%%%%%%

For this section we return to $\text{AB}_\text{out}\text{AB}_\text{in}$ for comparison against our previously reported results \cite{Kaulfuss_thesis:2021}. The net result of adding backscattering on the HOMM probability contour plot is a degradation of the edges of the plot. We have previously found that in linear chains of non-identical MRRs the effect of one of the MRRs in the chain having a larger (or smaller) round trip phase shift relative to the other is the shifting of the main crescent structure and the creation of a `spike structure' on the edge of the contour plot at either high or low $\theta$ values, near where the MRR is on resonance \cite{Kaulfuss_thesis:2021}. It is possible that backscattering can eliminate those spike structures and cause the parallel chains of MRRs to behave more like a chain of identical MRRs even if there are small differences in the design/production that cause small $\theta$ shifts in the chain. However, this also means that if one chooses to operate at these `spike structures' near MRR resonance, interference causes light to preferentially occupy cavity modes instead of waveguide modes and as a result the device is much more susceptible to the degradation of backscattering.

\begin{figure}[H]
    \centering
    \includegraphics[width=0.45\linewidth,keepaspectratio]{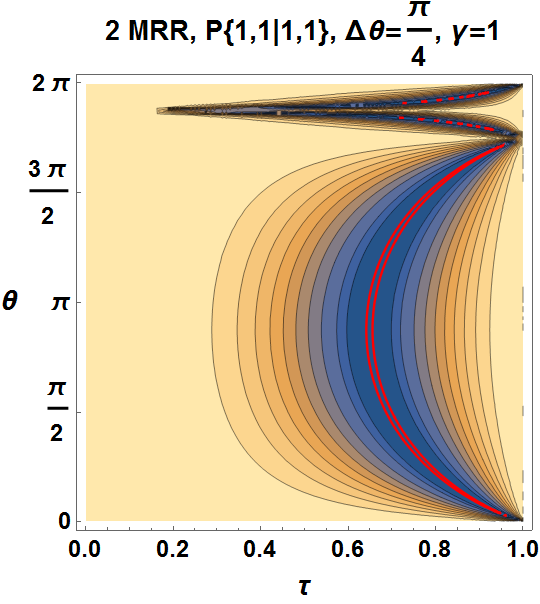} 
    \includegraphics[width=0.45\linewidth,keepaspectratio]{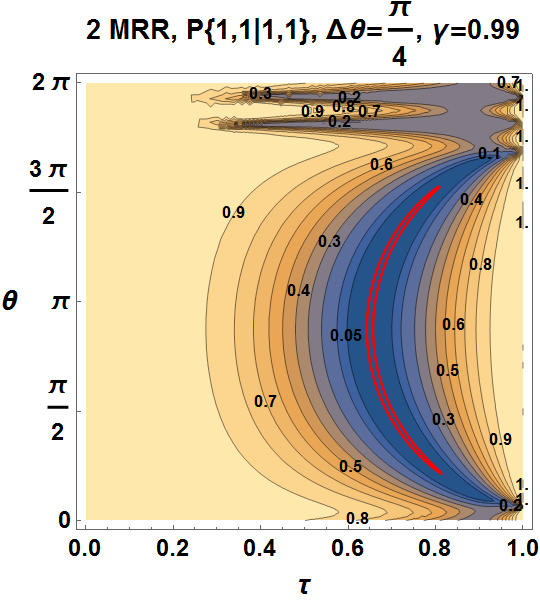} 
    \caption{2 MRR Non-Identical parallel, $\theta$ shift of $\frac{\pi}{4}$, with $\gamma=1$ (left) and $\gamma=0.99$ (right). Contour labels have been removed from the $\gamma=1$ plot, so the `spike' near $\theta=2\pi$ is visible. The red contour is a probability of $10^{-3}$.}
    \label{pi4g1}
\end{figure}

As an example we have included a 2 MRR parallel chain  with $\theta_1=\theta$ and $\theta_2=\theta+\frac{\pi}{4}$ in \Fig{pi4g1}. This is a larger phase shift than one would expect for fabrication differences, but it illuminates the effect of backscattering on the `spike structure'. As shown in \Fig{pi4g1}, an experimentally feasible value of  1\% backscattering ($\gamma=0.99$) destroys the HOMM in the `spike structure' and causes splitting.

However, these `spike structures' only exist at the edges of the contour plots, where the MRR is on resonance, far from the optimal operating regime of the HOMM at the center of the dark blue crescent, at $\theta=\pi$ and $\tau=\tau_c$. These small differences in round-trip phase shift between rings would never be noticed unless the device is operated near resonance, at $\theta=0$ or $\theta=2\pi$. If, for some reason, the device is to be operated at the edges of these contour plots, the inclusion of a small level of backscattering would cause any `spike' effects due to small differences between sequential MRRs to be removed and cause them to behave more like a chain of perfectly identical MRRs. 

\clearpage
\newpage
\section{Conclusion} \label{Conclusion}

From the results obtain from our model of backscattering, we can see that the HOMM is very robust against backscattering in MRRs, preserving the optimal operating region even in cases with extremely high backscattering, which would be uncharacteristic of a device fabricated within experimental tolerances. This is because the optimal operating region for the HOMM in MRRs is off resonance and the backscattering only has a destructive effect at $\theta$ values where the MRR is on resonance. Backscattering only deteriorates the areas where the MRR is on resonance, where you would not operate experimentally to achieve the HOMM, and therefore, we believe, these small levels of backscattering can be ignored when operating the MRR off resonance. Allowing MRRs to have a small level of backscattering does not destroy the HOMM in the optimal operating region ($\theta=\pi$ and $\tau=\tau_c$) and actually removes edge effects (on resonance) that occur due to slight differences between successive MRRs connected in a parallel chain. Overall, if an MRR-based device is operated off resonance to achieve the HOM effect the device is extremely robust against backscattering. However, if an MRR-based device is operated on resonance even a small amount of backscattering will completely destroy the HOM effect. 

We realize it appears somewhat counter-intuitive to operate an MRR completely ``off-resonance.'' However, the reason for doing so is because this is the optimal area to achieve the HOM effect. The main strength (or weakness depending on the application) of a resonator is its sensitivity to the environment; this is what allows resonators to make such good sensors. In our current investigation, where we are trying to obtain and maintain an interference effect, we want to be as stable as possible and therefore we want to operate the MRR not necessarily as a resonator, but as an extended beam splitter with more tunable parameters, which is what the MRR can be thought of when operated off resonance. The model we employed currently only considers backscattering from sidewall roughness inside the MRR. In future work, we plan to incorporate backscattering at the directional couplers 
(where external optical fibers couple to MRR device)
to better model what is happening in actual experiments. 

Lastly, the work considered here explored  the multidimensional solution space (HOMM) of the conventional HOM effect with a pair of photons input to the BS (one in each input port). Recently, some of the authors of this present work considered the interference effects for an arbitrary number of photons input into each port of a conventional BS. The conclusion of that work was that for any odd parity state (containing only combinations of odd number of photons) entering the A port of a 50:50, then for any state, pure or mixed, entering the B input port, there would be line of zeros along the output joint probability distribution for coincidence detection (i.e. the HOM effect would occur for any coincidence of an equal number of photons in each of the two output ports). They termed this effect the extended HOM effect (eHOM) \cite{eHOM_Alsing:2022,eHOM_Conf_Alsing:2022}. In future work, we will investigate further extending the eHOM effect by replacing the conventional BS with a MRR. 

%{\color{red} NOTE: this obviously raises the question: why not use an MZI?  Do we have a good response to this question? 
%\newline
%(PMA) Peter, this is a good point.  Possibly we'd get the same results using an extended MZI? If so, we could argue that the MRR is more compact, stable, \ldots, and therefore more useful for seeing the HOMM effect.} 
%%%%%%%%%%%%%%%%%%%%%%%%%%%%%%%%%%%%%%%%%%%%%%%%%%%%%%%%%%%%%%%%%%%%
%Remaining thoughts: Temporal stuff:
%I dont think backscattering has any impact on this so its a question of does the temporal stuff have an effect on the HOMM effect regardless of loss/backscattering
%Adding directional coupler reflections in addition to current interior backscattering would not change anything, just make the resonance splitting asymmetric... 
%Still always the question of why would you use an MRR off resonance instead of just an MZI. Perhaps increased stability?
%Not sure if this appendix stuff is worth mentioning, do people already know this? 

\begin{acknowledgments}
P.L.K. and R.J.B. would like to thank the Griffiss Institute (GI) for support of this work. P.M.A. would like to acknowledge support from the Air Force Office of Scientific Research (AFOSR). The authors wish to thank Christopher C. Tison for helpful discussions. Any opinions, findings and conclusions or recommendations expressed in this material are those of the author(s) and do not necessarily reflect the views of the Air Force Research Laboratory (AFRL) and the Army Research Laboratory (ARL).
\end{acknowledgments}

% References
\bibliography{references} % bibliography data in report.bib

\begin{thebibliography}{10}

\bibitem{Hach:2014}
Hach~III, E.~E., Preble, S.~F., Elshaari, A.~W., Alsing, P.~M., and Fanto,
  M.~L., ``Scalable {H}ong-{O}u-{M}andel manifolds in quantum-optical ring
  resonators,'' {\em Phys. Rev. A}~{\bf 89},  043805 (2014).

\bibitem{Alsing_Hach:2017a}
Alsing, P.~M., Hach~III, E.~E., Tison, C.~C., and Smith, A.~M., ``A quantum
  optical description of losses in ring resonators based on field operator
  transformations,'' {\em Phys. Rev. A}~{\bf 95},  053828 (2017).

\bibitem{Kaulfuss_thesis:2021}
Kaulfuss, P.~L.,  [{\em Using Linear Arrays of Micro-Ring Resonators to Enhance
  the {H}ong-{O}u-{M}andel Effect for Quantum Optical Networks in Silicon
  Nanophotonics}{\nolinebreak\hspace{0.1em}]}, School of Physics and Astronomy
  (COS), Rochester Institute of Technology (2021).

\bibitem{Kaulfuss:2023.Identical}
Kaulfuss, P.~L., Alsing, P.~M., Smith, A.~M., Monteleone, J., and Hach, E.~E.,
  ``Enhanced hong-ou-mandel manifolds and figures of merit for linear chains of
  identical microring resonators,'' {\em Phys. Rev. Res.}~{\bf 5},  023097 (May
  2023).

\bibitem{eHOM_Alsing:2022}
Alsing, P.~M., Birrittella, R.~J., Gerry, C.~C., Mimih, J., and Knight, P.~L.,
  ``Extending the {H}ong-{O}u-{M}andel effect: The power of nonclassicality,''
  {\em Phys. Rev. A}~{\bf 105},  013712 (Jan 2022).

\bibitem{eHOM_Conf_Alsing:2022}
Alsing, P.~M., Birrittella, R.~J., Gerry, C.~C., Mimih, J., and Knight, P.~L.,
  ``The {H}ong-{O}u-{M}andel effect is really odd,'' in [{\em Quantum 2.0
  Conference and Exhibition}{\nolinebreak\hspace{0.1em}]},  {\em Quantum 2.0
  Conference and Exhibition}~{\bf 1},  QW3A.37, Optica Publishing Group (2022).

\bibitem{HOM:1987}
Hong, C., Ou, Z., and Mandel, L., ``Measurement of subpicosecond time intervals
  between two photons by interference,'' {\em Phys. Rev. Lett.}~{\bf 59},  2044
  (1987).

\bibitem{Bouchard_2021}
Bouchard, F., Sit, A., Zhang, Y., Fickler, R., Miatto, F.~M., Yao, Y.,
  Sciarrino, F., and Karimi, E., ``Two-photon interference: the
  {H}ong{\textendash}{O}u{\textendash}{M}andel effect,'' {\em Reports on
  Progress in Physics}~{\bf 84},  012402 (jan 2021).

\bibitem{MRRReview:Bogaerts:2012}
Bogaerts, W., De~Heyn, P., Van~Vaerenbergh, T., De~Vos, K., Kumar~Selvaraja,
  S., Claes, T., Dumon, P., Bienstman, P., Van~Thourhout, D., and Baets, R.,
  ``Silicon microring resonators,'' {\em Laser \& Photonics Reviews}~{\bf
  6}(1),  47--73 (2012).

\bibitem{BackscatterReview:Li:2016}
Li, A., Van~Vaerenbergh, T., De~Heyn, P., Bienstman, P., and Bogaerts, W.,
  ``Backscattering in silicon microring resonators: a quantitative analysis,''
  {\em Laser \& Photonics Reviews}~{\bf 10}(3),  420--431 (2016).

\bibitem{Hance:2021}
Hance, J.~R., Sinclair, G.~F., and Rarity, J., ``Backscatter and spontaneous
  four-wave mixing in micro-ring resonators,'' {\em Journal of Physics:
  Photonics}~{\bf 3},  025003 (apr 2021).

\bibitem{Ballesteros:2011}
Ballesteros, G.~C., Matres, J., Mart\'{i}, J., and Oton, C.~J.,
  ``Characterizing and modeling backscattering in silicon microring
  resonators,'' {\em Opt. Express}~{\bf 19},  24980--24985 (Dec 2011).

\bibitem{McCutcheon:2021}
McCutcheon, W., ``Backscattering in nonlinear microring resonators via a
  gaussian treatment of coupled cavity modes,'' {\em APL Photonics}~{\bf 6}(6),
   066103 (2021).

\bibitem{Little:1997}
Little, B., Chu, S., Haus, H., Foresi, J., and Laine, J.-P., ``Microring
  resonator channel dropping filters,'' {\em Journal of Lightwave
  Technology}~{\bf 15}(6),  998--1005 (1997).

\bibitem{tau:real:note}
If one were to let the coupling have phase $\tau=|\tau|\,e^{i\,\varphi}$, the
  only change to the MRR transfer matrix elements would be that the internal
  phase accumulation angle $\theta$ would be modified to $\theta\to\theta' =
  \theta+\varphi$. \cite{Rabus:2007}. While this would shift the location of
  the HOM maximum away from $\theta=\pi$ (as discussed in the main text), the
  transmission curves would also shift commensurately, so that the maximum HOM
  would still occur halfway between the free spectral range of the MRR, as for
  $\tau$ real. Thus, one may, without loss of generality, take $\tau$ to be
  real.

\bibitem{Skaar:2004}
Skaar, J., Escartin, J., and Landro, H., ``Quantum mechanical description of
  linear optics,'' {\em Am. J. Phys.}~{\bf 72},  1385 (2004).

\bibitem{aym:splitting:note}
Asymmetric splitting can be achieved by employing another set of beam splitters
  (BS) on the external busses to the MRR. For simplification of the model, and
  without loss of generality, we have chosen not to include such additional BS.

\bibitem{Yariv:2000}
Yariv, A., ``Universal relations for coupling of optical power between
  microresonators and dielectric waveguides,'' {\em Electronic Letts.}~{\bf
  36},  321 (2000).

\bibitem{Rabus:2007}
Rabus, D.~G.,  [{\em Integrated Ring Resonators}{\nolinebreak\hspace{0.1em}]},
  Springer-Verlag, Berlin (2007).

\end{thebibliography}
\bibliographystyle{spiebib} % makes bibtex use spiebib.bst

%=======================================================================
% Appendices
%=======================================================================
\clearpage
\newpage
%=======================================================================
\appendix
%=======================================================================
%\clearpage
%\newpage

\section[Appendix A]{Interpretation of MRR input/output \\ transfer matrix}\label{Appendix:1}
\bigskip

\begin{equation}
    \begin{pmatrix}
    \hat{a}_{in}^\dagger \\
    \hat{b}_{in}^\dagger \\
    \hat{c}_{in}^\dagger \\
    \hat{d}_{in}^\dagger
    \end{pmatrix}
    =
    \begin{pmatrix}
    A_1 & B_1 & C_1 & D_1 \\
    A_2 & B_2 & C_2 & D_2 \\
    A_3 & B_3 & C_3 & D_3 \\
    A_4 & B_4 & C_4 & D_4 
    \end{pmatrix}
    \begin{pmatrix}
    \hat{a}_{out}^\dagger \\
    \hat{b}_{out}^\dagger \\
    \hat{c}_{out}^\dagger \\
    \hat{d}_{out}^\dagger
    \end{pmatrix}
    \label{generalmatrixeqAPPENDIX}
\end{equation}

We take this opportunity to illuminate the power of the matrix element naming scheme we have chosen for the MRR input/output transfer matrix $\mathcal{S}$ in \Eq{generalmatrixeqAPPENDIX}. Notice that all entries in the same row all contain the same subscript index, and all entries in the same column all contain the same alphabetic letter. This allows us to quickly transcribe the conditional probability for a set of given inputs and outputs. We can do this because now the Fock state input determines the subscript number(s) that must be present in each term of the probability and then the output state determines the letter(s). To obtain the full probability we them simply sum up all combinations of distributing the possible subscripts across the letters present in each term. 

As an example: using the general $4\times 4$ transfer matrix $\mathcal{S}$ above in \Eq{generalmatrixeqAPPENDIX}, 
let the output state be given by $\ket{1,2,0,2}_{out}$. Looking down columns, this tells us that all of our probability terms will take the form $ABBDD$, that is $1\, A, 2\, Bs, 0\, Cs, 2\, Ds$. The input state, say $\ket{1,3,1,0}_{in}$, tells us what subscript numbers we can distribute amongst the previous letters, 
in this case: 1 1-subscript, 3 2-subscripts, 1 3-subscript and 0 4-subscript. Therefore, in this example, the first term in the conditional probability would be $A_1B_2B_2D_2D_3$, and all other terms could be constructed by summing all possible ways to distribute one `1', three `2's, and one `3' across the letters $ABBDD$, that is
$P = |A_1B_2B_2D_2D_3 + A_2B_1B_2D_2D_3 + A_2B_2B_1D_2D_3 + \ldots|^2$.

As a relevant example, the AB input state $\ket{1,1,0,0}_{in}$ indicates that there must be 1 1-subscript and 1-2-subscript. If we look for the AB output state $\ket{1,1,0,0}_{out}$, then the output must be of the form $AB$. Distributing the allowable subscripts over all letters yields the probability amplitude $A_1 B_2+A_2 B_1$, which is the permanent of the upper left-hand corner of $\mathcal{S}$. The HOMM probability condition is then 
$P=|A_1 B_2+A_2 B_1|^2=0$, which is \Eq{PABAB}.
If we instead were to consider coincidence on the $CD$ output modes i.e. $\ket{0,0,1,1}_{out}$, the only thing that would change is that the above mnemonic would indicate that $AB\to CD$. Thus, the HOMM probability condition would then instead be $P=|C_1 D_2+C_2 D_1|^2=0$, involving the square of the permanent of the upper right-hand corner of $\mathcal{S}$.
This was used in \Sec{CDoutABin}.
Finally, in \Sec{ADoutADin} we considered the $AD$ input state $\ket{1,0,0,1}_{in}$ indicating 1 1-index and 1 4-index. We considered the $AD$ output state $\ket{1,0,0,1}_{out}$ indicating that output terms of the form $AD$. This then yields the HOMM probability condition 
$P=|A_1 D_4+A_4 D_1|^2=0$, which is the permanent of the outer four corner terms of $\mathcal{S}$.

The conventional HOM effect using a standard 50:50 (balanced) beam splitter (BS) occurs because the amplitude for the output coincident state $\ket{1,1}_{AB}$ contains two terms: one for both photons transmitting through the BS, and the other for both photons reflecting off the BS. For a 50:50 BS, both amplitudes have the same magnitude, but opposite sign, and hence cancel, yielding a zero output probability. 
For the BS transfer matrix Eqn.(\ref{BSmatrix})
$\tiny{U_{BS}=\left(
\begin{array}{cc}
\tau & \kappa \\
-\kappa & \tau
\end{array}
\right)}$
the HOM condition occurs at the condition Perm$(U_{BS})=0$.
Defining $\tau=\cos(\varphi/2)$ and $\kappa=\sin(\varphi/2)$ for $0\le \varphi\le \pi$, the above condition occurs for $\cos(\varphi)=0 \Rightarrow \varphi = \pi/2$, i.e. the 50:50 BS condition ($\tau=\kappa$).
When we generalize from a conventional BS to a MRR (or parallel array of MRRs), the HOM condition continues to be defined by the zero value of the permanent of the relevant transfer matrix, as discussed above. Since each MRR contains 3 parameters $\{\tau_{upper}, \tau_{lower}, \theta\}$, the  transmission amplitudes for the upper and lower BS coupling junctions (into/out of the MRR from the waveguide bus) and the MRR round trip phase accumulation angle, the Perm$(U_{MRR})=0$ condition gives rise to a multidimensional manifold of (parameter) solutions, vs the single point (0D manifold, $\theta=\pi/2$) solution of the conventional BS.

\clearpage
\newpage

\begin{comment}
\section{Loss extra example plots}\label{Appendix:2}

{\color{red} New loss plots to think about:}

\begin{figure}[H]
    \centering
    \includegraphics[width=0.75\linewidth,keepaspectratio]{probg=1a=0.99.png} 
    \includegraphics[width=0.75\linewidth,keepaspectratio]{probg=1a=0.95.png} 
\end{figure}

\begin{figure}[H]
    \centering
    \begin{tabular}{cc}
    \includegraphics[width=0.45\linewidth,keepaspectratio]{contourg=1a=0.99.png} &
    \includegraphics[width=0.45\linewidth,keepaspectratio]{contourg=1a=0.95.png}
    \end{tabular}
    \caption{some loss plots}
    \label{somelossplots1}
\end{figure}
\begin{figure}[H]
    \centering
    \includegraphics[width=0.75\linewidth,keepaspectratio]{probg=0.95a=0.99.png}
    \includegraphics[width=0.75\linewidth,keepaspectratio]{probg=0.95a=0.95.png} 
\end{figure}

\begin{figure}[H]
    \centering
    \begin{tabular}{cc}
    \includegraphics[width=0.5\linewidth,keepaspectratio]{contourg=0.95a=0.99.png} & \includegraphics[width=0.5\linewidth,keepaspectratio]{contourg=0.95a=0.95.png}
    \end{tabular}
    \caption{some loss plots}
    \label{somelossplots2}
\end{figure}
\end{comment}

%%%%%%%%%%%%%%%%%%%%%%%%%%%%%%%%%%%%%%%%%%%%%%%%%%%%%%%%%%%%%%%%%%%%
\end{document}